\pdfoutput=1

\documentclass{raa}

\usepackage{mathptmx}
\usepackage{ae,aecompl}

\usepackage{graphicx}	
\usepackage{amsmath}	
\usepackage{amssymb}	
\usepackage{pifont}
\usepackage{txfonts}
\usepackage{url}
\usepackage{subfigure}
\usepackage{longtable}
\usepackage{lscape}
\usepackage{multirow}
\usepackage{color}
\usepackage{textcomp}
\usepackage{natbib}
\usepackage[colorlinks,linkcolor=red,anchorcolor=green,citecolor=blue]{hyperref}

\newcommand{\HI}{H\,{\small{I}} }

\begin{document}

\title{Radio Frequency Interference Mitigation and Statistics in the Spectral Observations of FAST}


   \volnopage{Vol.0 (2021) No.0, 000--000}      
   \setcounter{page}{1}          
   \author{Chuan-Peng Zhang\inst{1,2},
      Jin-Long Xu\inst{1,2},
      Jie Wang\inst{1},
      Yingjie Jing\inst{1},
      Ziming Liu\inst{1,3},
      Ming Zhu\inst{1,2},
      Peng Jiang\inst{1,2}
   }

   \institute{National Astronomical Observatories, Chinese Academy of Sciences, 100101 Beijing, China; {\it cpzhang@nao.cas.cn} \\
    \and
    CAS Key Laboratory of FAST, NAOC, Chinese Academy of Sciences, 100101 Beijing, China \\
    \and
    College of Astronomy and Space Sciences, University of Chinese Academy of Sciences, 100049 Beijing, China
   }


\abstract{In radio astronomy, radio frequency interference (RFI) becomes more and more serious for radio observational facilities. The RFI always influences the search and study of the interesting astronomical objects. Mitigating the RFI becomes an essential procedure in any survey data processing. Five-hundred-meter Aperture Spherical radio Telescope (FAST) is an extremely sensitive radio telescope. It is necessary to find out an effective and precise RFI mitigation method for FAST data processing. In this work, we introduce a method to mitigate the RFI in FAST spectral observation and make a statistics for the RFI using $\sim$300 hours FAST data. The details are as follows. Firstly, according to the characteristics of FAST spectra, we propose to use the ArPLS algorithm for baseline fitting. Our test results show that it has a good performance. Secondly, we flag the RFI with four strategies, which are to flag extremely strong RFI, flag long-lasting RFI, flag polarized RFI, and flag beam-combined RFI, respectively. The test results show that all the RFI above a preset threshold could be flagged. Thirdly, we make a statistics for the probabilities of polarized \texttt{XX} and \texttt{YY} RFI in FAST observations. The statistical results could tell us which frequencies are relatively quiescent. With such statistical data, we are able to avoid using such frequencies in our spectral observations. Finally, based on the $\sim$300 hours FAST data, we got an RFI table, which is the most complete database currently for FAST. 
\keywords{radio telescope: FAST --- radio frequency interference --- method --- RFI mitigation} }

   \authorrunning{Chuan-Peng Zhang, et al.}                     
   \titlerunning{Radio Frequency Interference Mitigation and Statistics}       

   \maketitle


\section{Introduction}    
\label{sect:intro}

In radio astronomy, radio frequency interference (RFI) becomes more and more serious for radio observational facilities \citep{Kesteven2005,An2017,Zeng2021}. The days of interference-free observations in radio astronomy are long gone. Substantial pressures are coming from commercial, defense, and other interests for greater access to the radio-frequency spectrum. This means that radio astronomers can no longer rely on the regulatory authorities for an environment free from interference, and must look seriously at mitigation strategies \citep{Fridman2001,Boonstra2005,Baan2011}.

Mitigating the RFI becomes an essential procedure in pulsar survey and other spectral data processing. RFI can occur in impulse type bursts persistently, variable in time or in a transient manner. The RFI properties in the integrated post-correlated data can be studied in terms of their morphology, consisting of broad band, narrow band roughly horizontal and vertical envelopes, mainly caused by the earlier described terrestrial sources, and unknown blob RFI \citep{An2017,Zeng2021}. The RFI amplitudes are typically much higher than the underlying astronomical signal and random noise, but some weak RFI signals are almost identical to the intensity of real celestial signals, such as pulsars and molecular spectra \citep{An2017,Zhang2020}. This will easily lead to misidentification and seriously influence the searching and analysis of the interesting astronomical objects. An important distinguishing feature for the separation of terrestrial sources of RFI from the astronomical signal relates to the amount of dispersion in the ``time-frequency'' domain of the received data \citep{An2017}.

FAST is located at a site position of the protected radio quiet zone in Guizhou of China. However, some strong RFI from for example, broadcast radio and TV, communication satellites and navigation satellites are extremely hard to be shielded. In addition, FAST is very susceptible to RFI from the electronic instruments integrated with the telescope and active radio services around the FAST site, due to its high sensitivity \citep{Zhang2020}. Therefore, we must look for another strategies to mitigate the RFI in FAST observations.

The aim of this work is to flag and mitigate the RFI from the FAST observations, and then give a probability and statistics for the FAST RFI showing up. Section\,\ref{sect:rfi_intro} introduces the features of astronomical emission lines and terrestrial RFI in FAST observations. Section\,\ref{sect:results} presents our proposed scheme for RFI mitigation in FAST spectral observations. Section\,\ref{sect:statis} shows the RFI statistical results using our proposed strategies. Section\,\ref{sect:summary} gives a summary.

\begin{figure*}
\centering
\includegraphics[width=0.99\textwidth, angle=0]{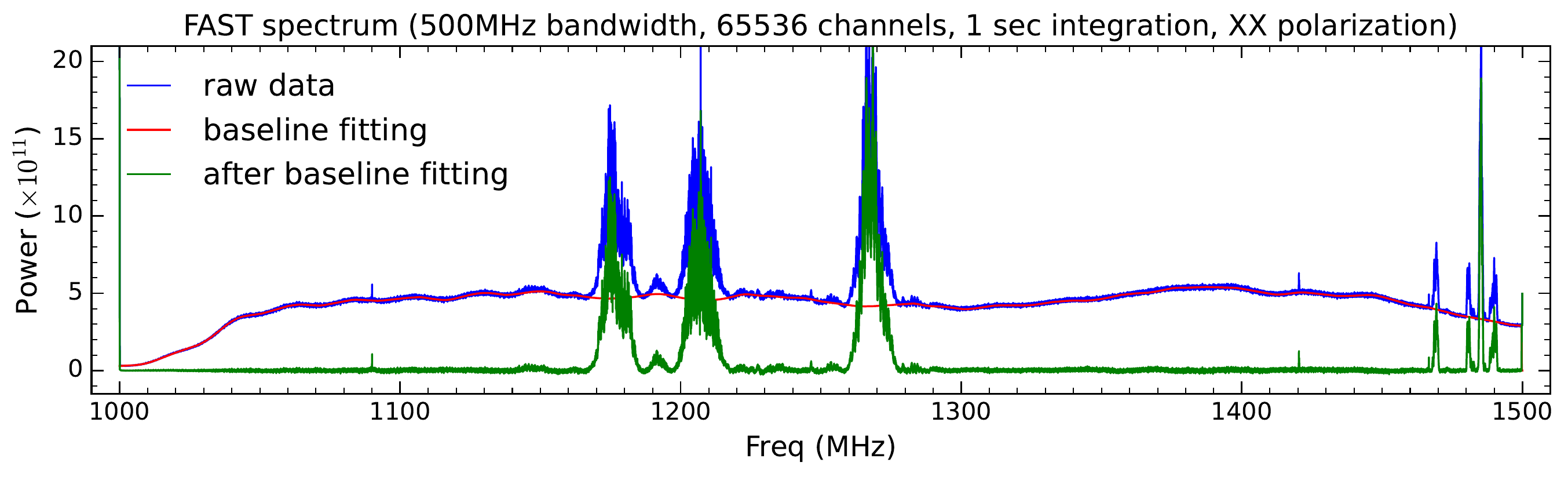}
\includegraphics[width=0.99\textwidth, angle=0]{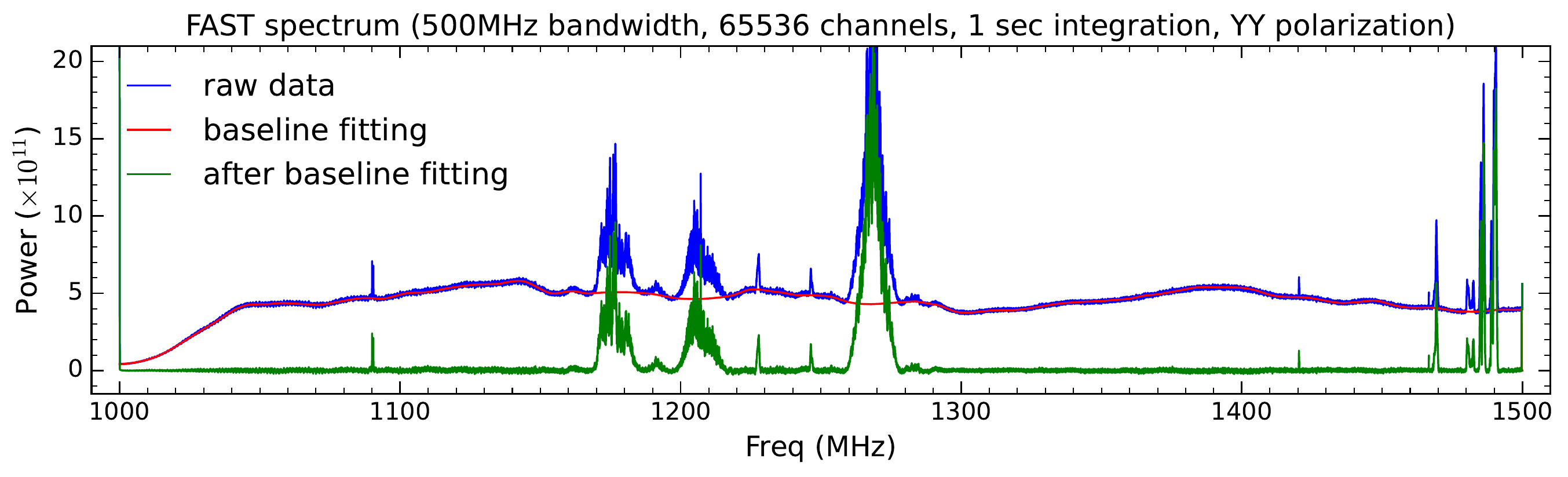}
\caption{Two full wideband of FAST spectra (polarizations: \texttt{XX} and \texttt{YY}; integration time: 1\,second) covering a frequency range from 1000 to 1500\,MHz. All the visible emission lines except \HI at $\sim$1420.5\,MHz are RFI from terrestrial sources. }
\label{Fig:baseline_fit}
\end{figure*}

\section{Astronomical emission lines and terrestrial RFI in FAST}
\label{sect:rfi_intro}

\subsection{The characteristics for the spectral data}
\label{sect:data}

FAST has a 19-beam receiver for simultaneously covering multiple observational areas, thus it is a good tool to carry out drift scan observations for searching for extra-galactic point sources. The data are recorded in the spectral-line backend using a dual linear polarization (\texttt{XX} and \texttt{YY}) mode with a relatively high time-resolution. The full wideband (500\,MHz) in the spectral-line backend has a frequency coverage from 1000 to 1500\,MHz (see Figure\,\ref{Fig:baseline_fit}). The high and low frequency resolution modes have a number of 1\,M and 64\,k channels, respectively. The frequency resolution for the high and low resolution modes is 476.84\,Hz and 7630\,Hz, respectively. The sampling time is 1\,second for each spectral line in this work. Generally, 2048\,seconds observation spectra for the 64\,k channels are packaged into a data cube that has a size of around 2.1\,GB. The aperture of the FAST is 500\,m and the effective aperture is about 300\,m corresponding to an HPBW of 2.9$'$ at 1.4\,GHz \citep{Nan2011,Jiang2019,Jiang2020,Qian2020,Han2021}.

In Figure\,\ref{Fig:baseline_fit}, we present a FAST 500\,MHz-bandwidth spectrum with 1\,second integration time. Each spectrum in the mode of low frequency resolution has 65536 (64\,k) channels in total. This leads to that the computing resources are time-consuming in reducing the FAST spectral data. We could firstly choose one out of every 100 channels, and then fit the baseline only for the selected channels. Finally, we could use interpolation technique to promote and estimate the baseline for the whole 65536 channels.

\subsection{The general RFI distribution}

Figure\,\ref{Fig:baseline_fit} shows a full wideband of FAST spectrum covering a frequency range from 1000 to 1500\,MHz. All the visible emission lines except \HI at $\sim$1420.5\,MHz in Figure\,\ref{Fig:baseline_fit} are RFI that comes from terrestrial sources. The RFI, including broadcast radio and TV, cell phones, communication satellites, navigation satellites, as well as all the wireless control and monitoring systems \citep{Zhang2020,Wang2021}, have the potential to affect radio-astronomical observations. The RFI pollution makes the background against what radio astronomers concern be noisy \citep{Boonstra2005}.

\section{The proposed scheme for RFI Mitigation}
\label{sect:results}

\begin{figure*}
\centering
\includegraphics[width=0.75\textwidth, angle=0]{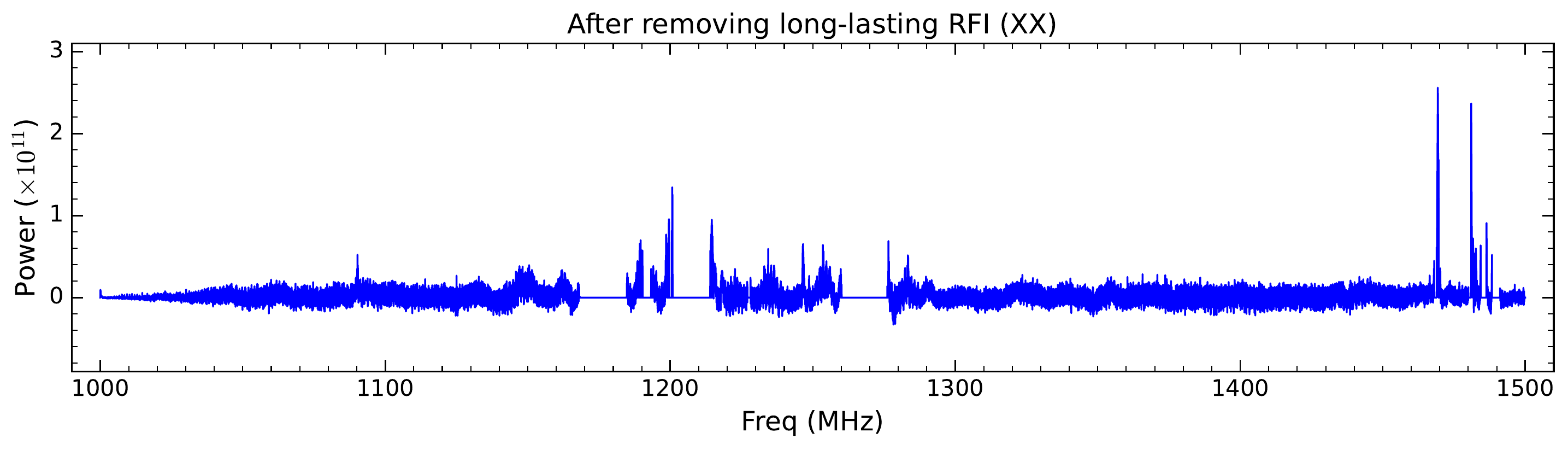}
\includegraphics[width=0.75\textwidth, angle=0]{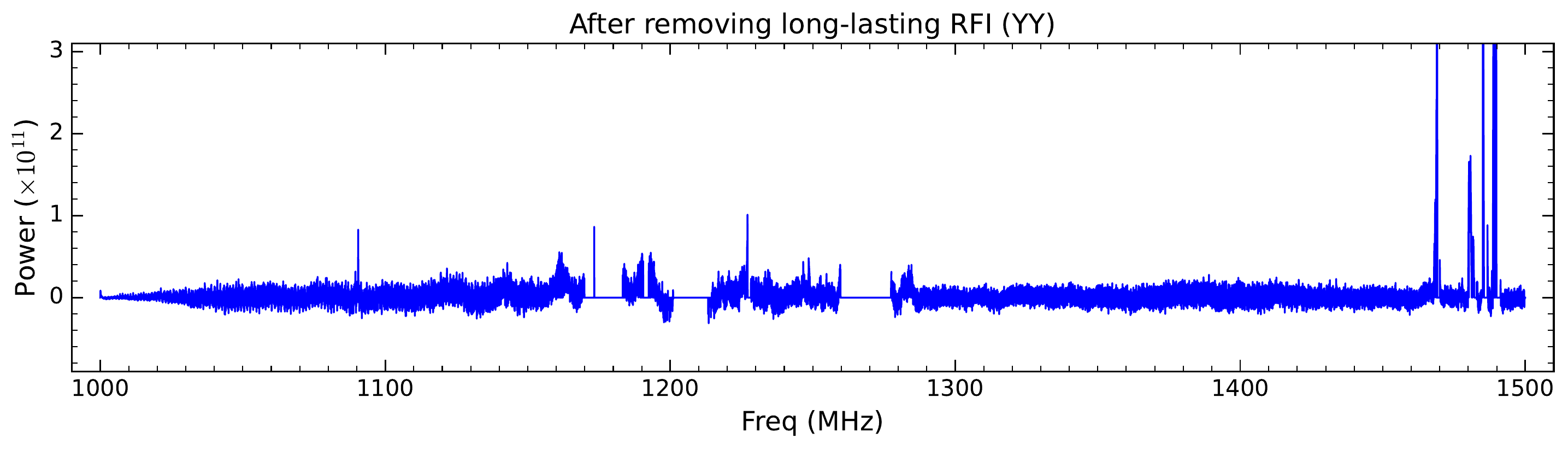}
\includegraphics[width=0.80\textwidth, angle=0]{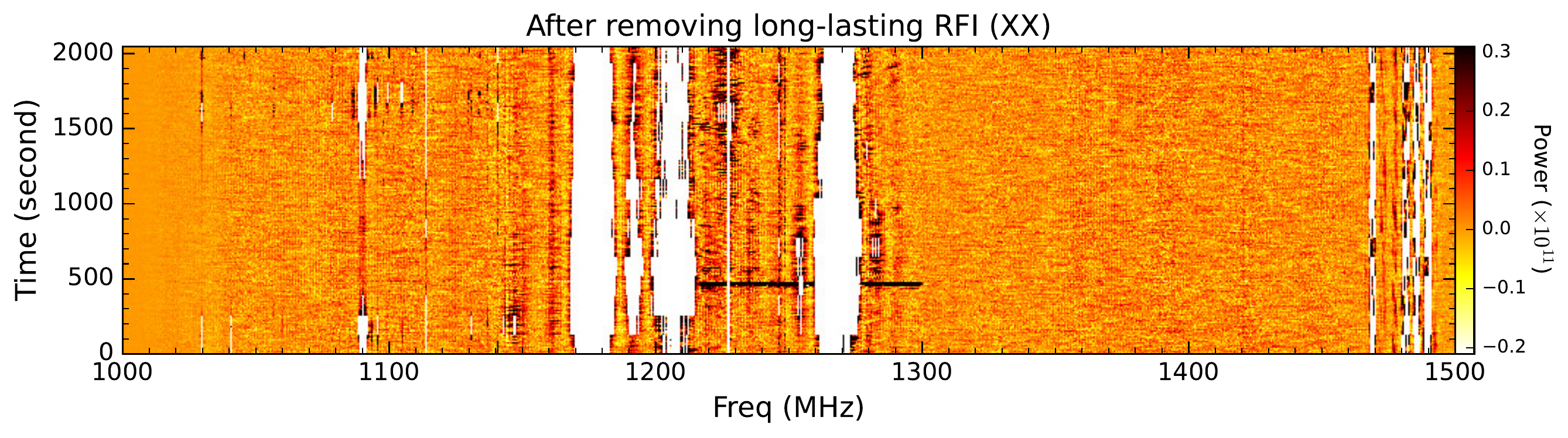}
\includegraphics[width=0.80\textwidth, angle=0]{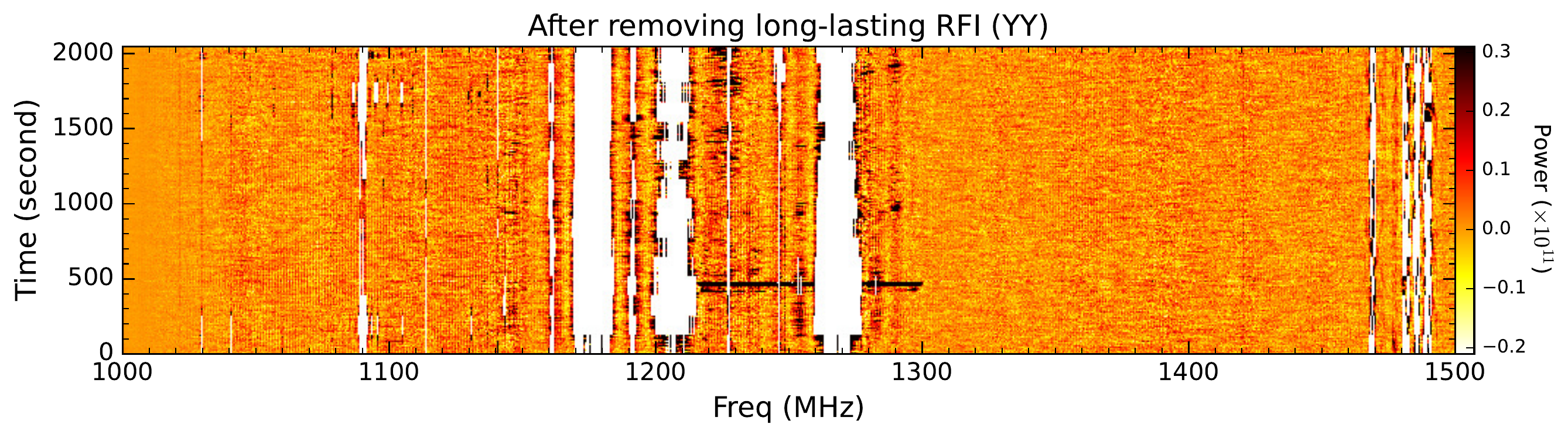}
\caption{Upper two panels: two FAST spectra (polarization: \texttt{XX} and \texttt{YY}; integration time: 1\,second) after removing long-lasting (> 128 seconds) RFI above a threshold of $3\sigma$ ($\sigma \approx 108{\rm\,mK}$). Lower two panels: two 2048 seconds waterfall images after removing the long-lasting RFI.}
\label{Fig:rfi_time}
\end{figure*}

\begin{figure*}
\centering
\includegraphics[width=0.75\textwidth, angle=0]{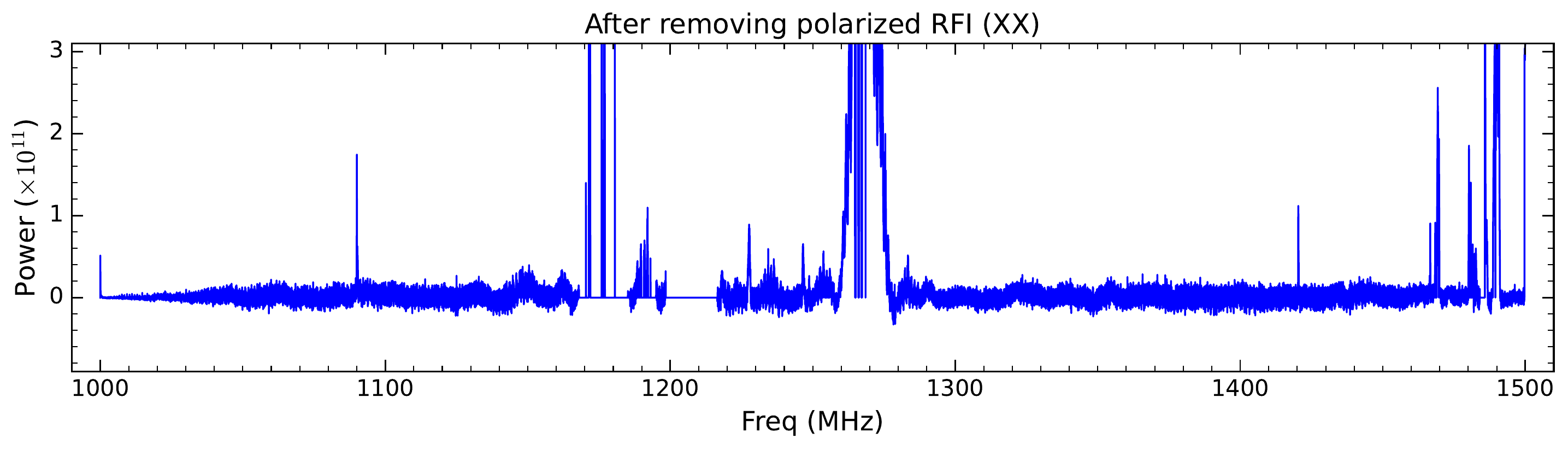}
\includegraphics[width=0.75\textwidth, angle=0]{./figures/ms2021-0304-fig07-eps-converted-to.pdf}
\includegraphics[width=0.80\textwidth, angle=0]{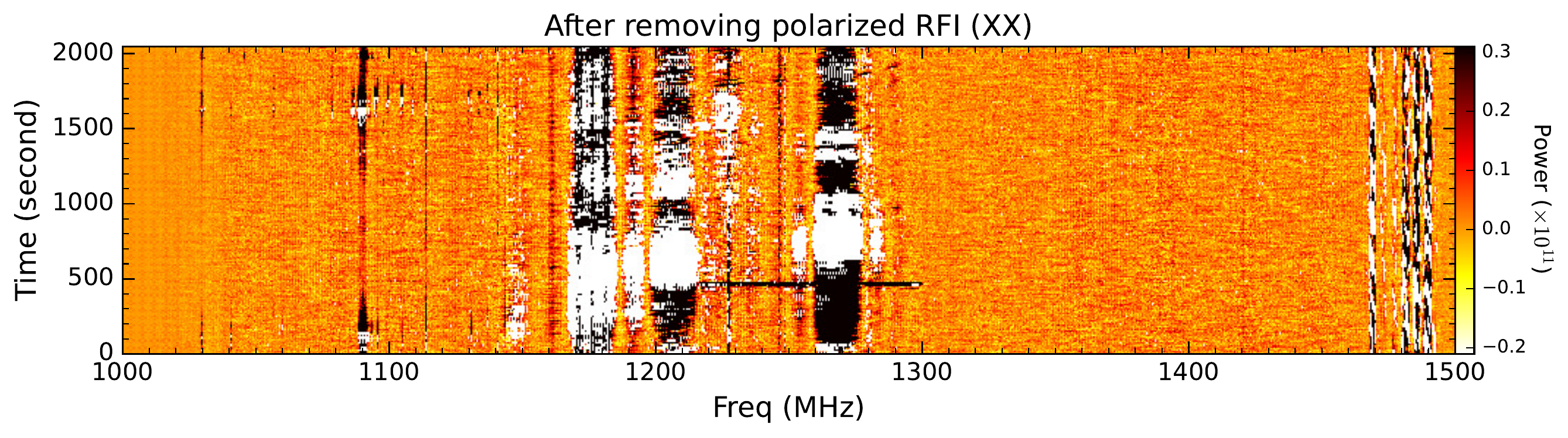}
\includegraphics[width=0.80\textwidth, angle=0]{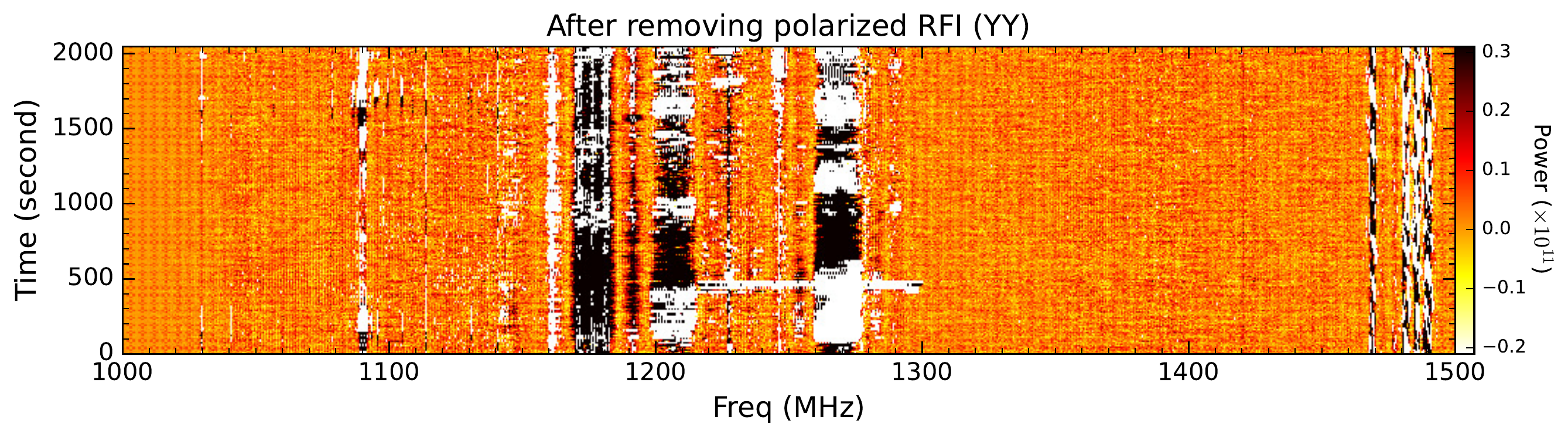}
\caption{Upper two panels: two FAST spectra (polarization: \texttt{XX} and \texttt{YY}; integration time: 1\,second) after removing polarized RFI above a threshold of $3\sigma$ ($\sigma \approx 108{\rm\,mK}$). Lower two panels: two 2048 seconds waterfall images after removing the polarized RFI.}
\label{Fig:rfi_pol}
\end{figure*}

\begin{figure*}
\centering
\includegraphics[width=0.75\textwidth, angle=0]{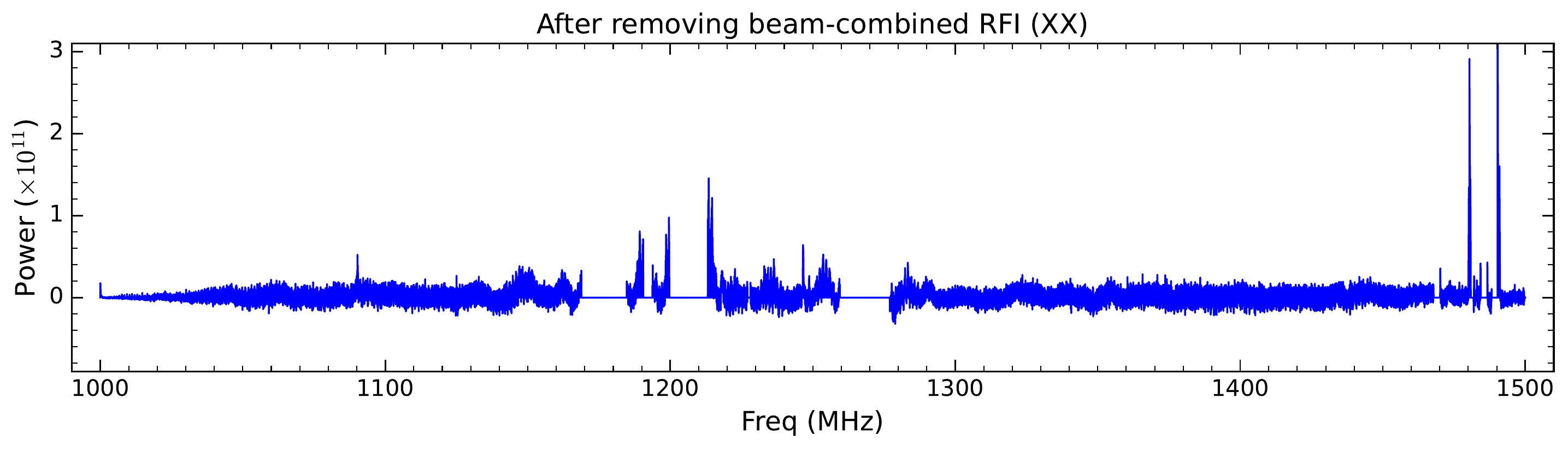}
\includegraphics[width=0.75\textwidth, angle=0]{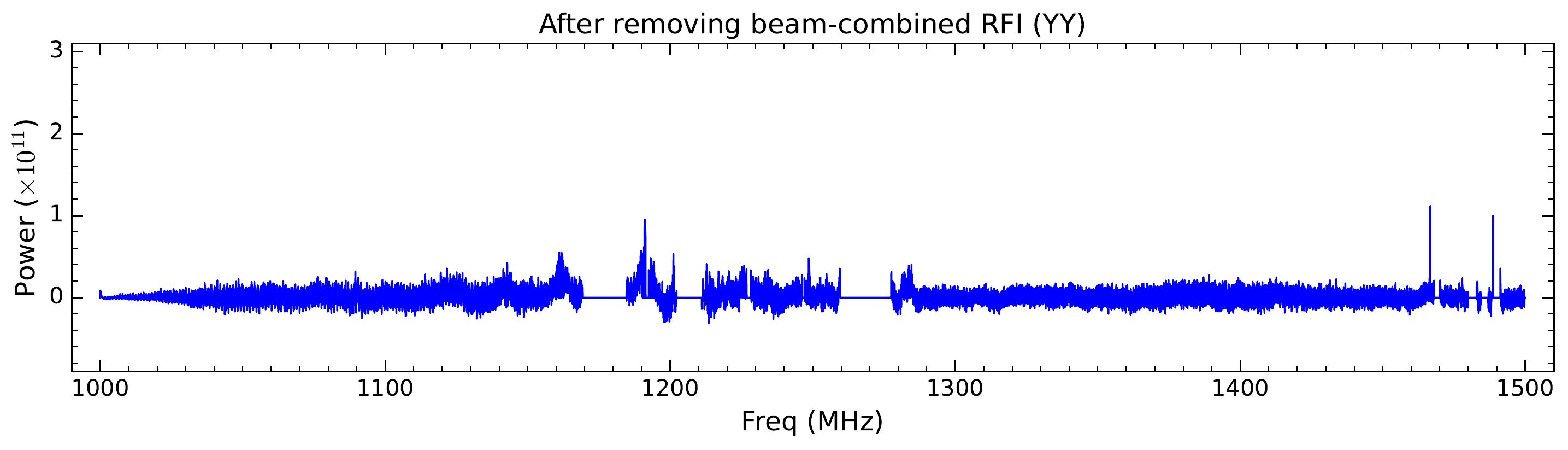}
\includegraphics[width=0.80\textwidth, angle=0]{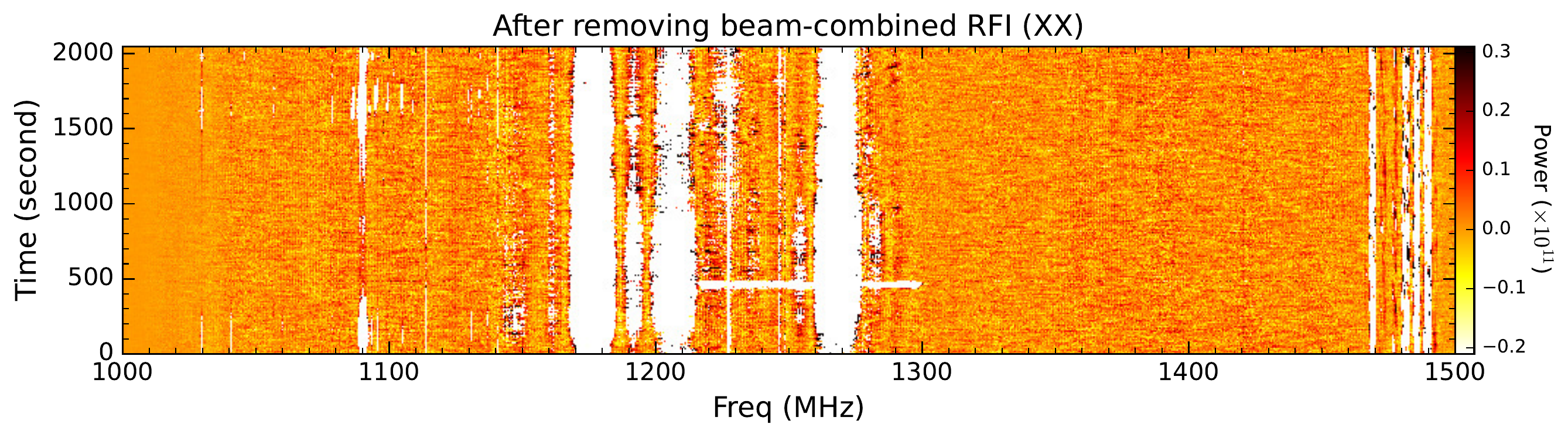}
\includegraphics[width=0.80\textwidth, angle=0]{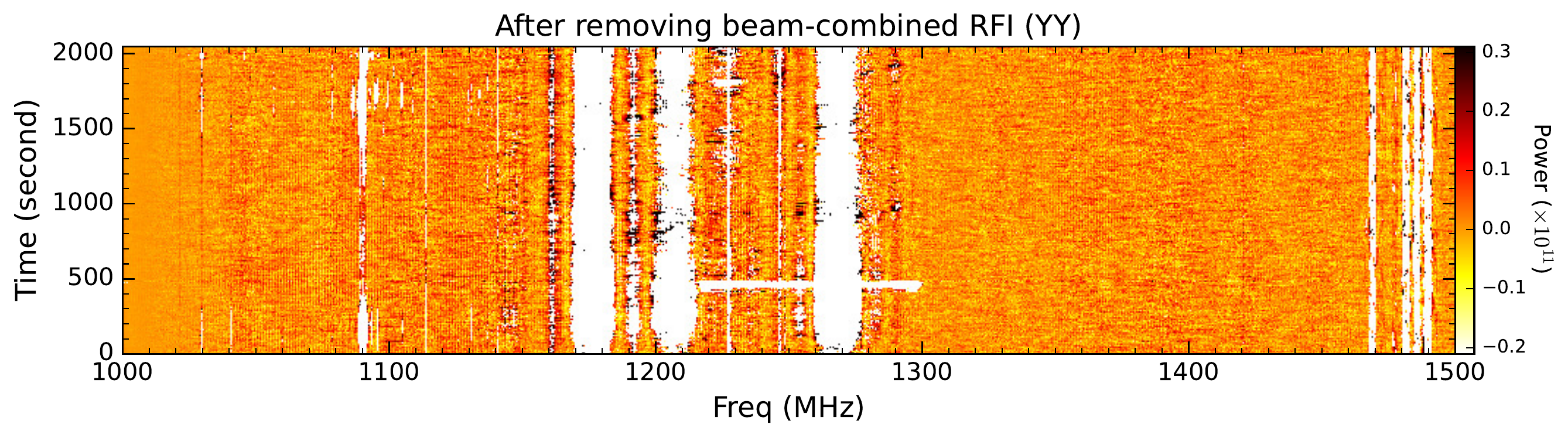}
\caption{Upper two panels: two FAST spectra (polarization: \texttt{XX} and \texttt{YY}; integration time: 1\,second) after removing beam-combined RFI above a threshold of $3\sigma$ ($\sigma \approx 108{\rm\,mK}$). Lower two panels: two 2048 seconds waterfall images after removing the beam-combined RFI.}
\label{Fig:rfi_beam}
\end{figure*}

\begin{figure*}
\centering
\includegraphics[width=0.75\textwidth, angle=0]{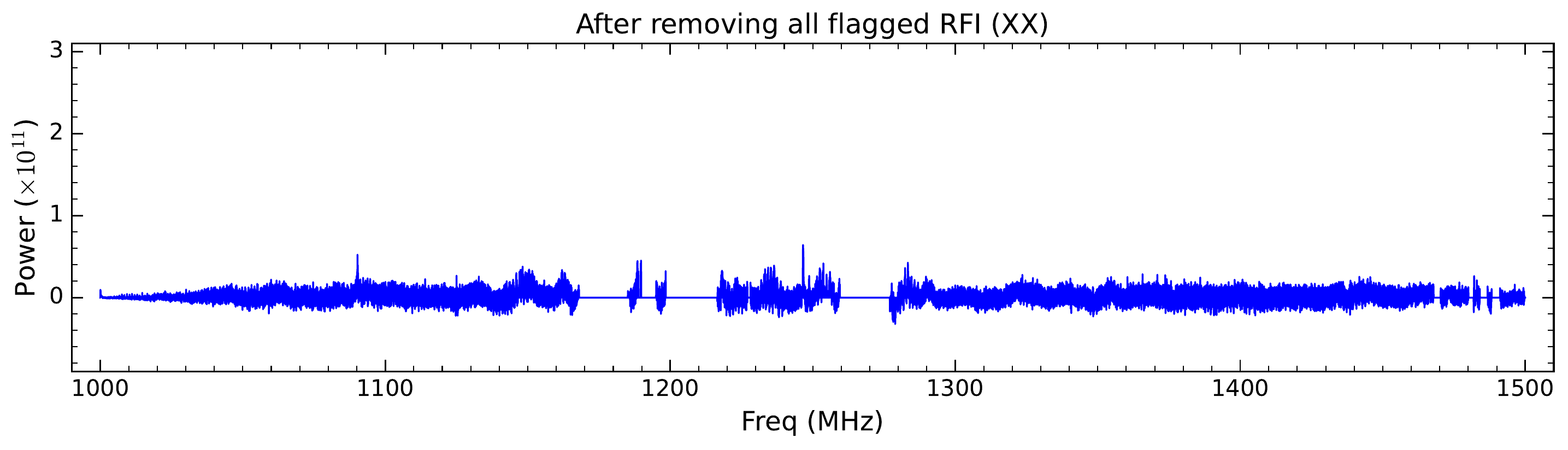}
\includegraphics[width=0.75\textwidth, angle=0]{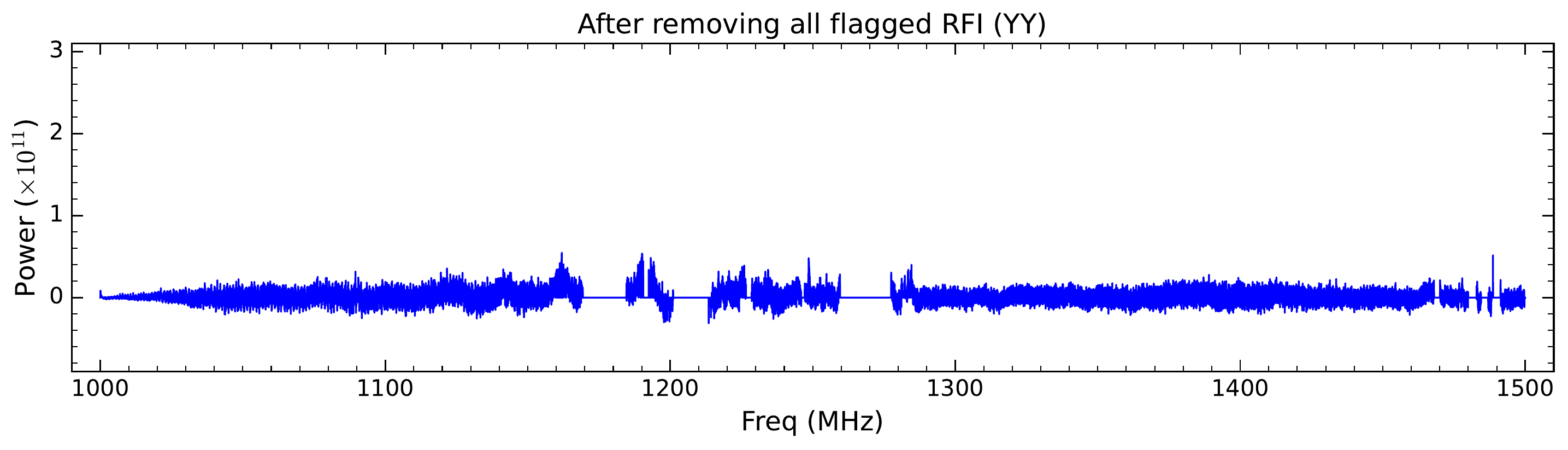}
\includegraphics[width=0.80\textwidth, angle=0]{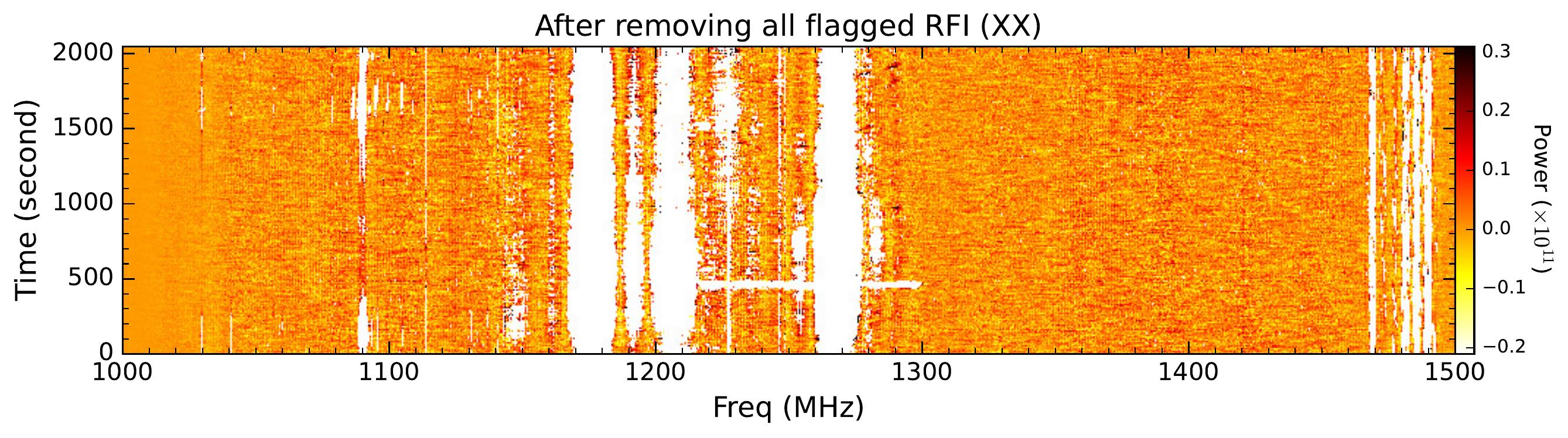}
\includegraphics[width=0.80\textwidth, angle=0]{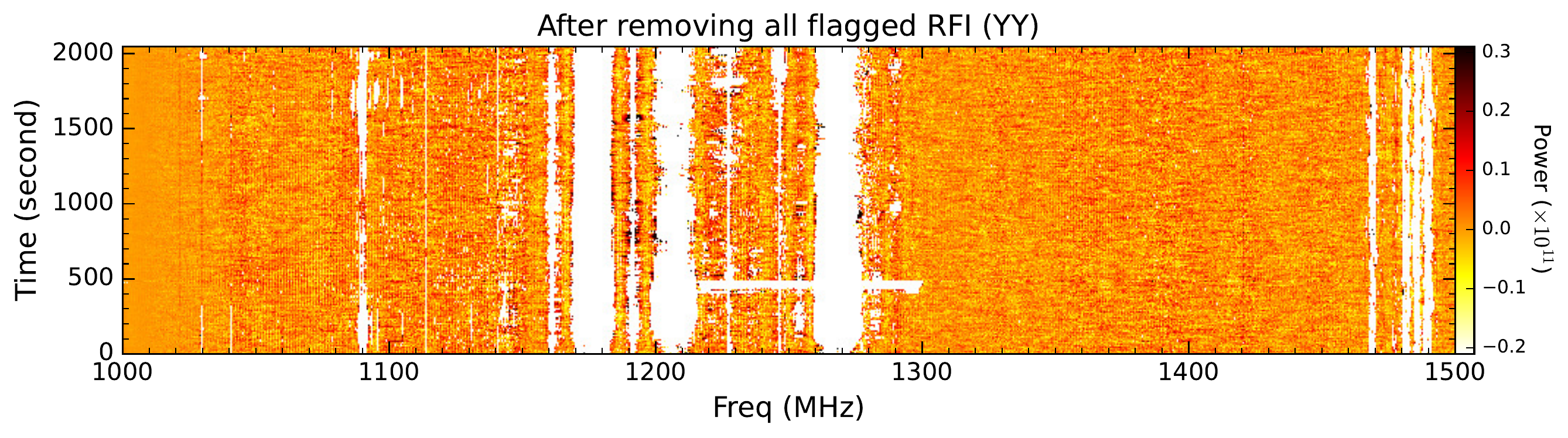}
\caption{Upper two panels: two FAST spectra (polarization: \texttt{XX} and \texttt{YY}; integration time: 1\,second) after removing all the flagged RFI. Lower two panels: two 2048 seconds waterfall images after removing all the flagged RFI.}
\label{Fig:rfi_all}
\end{figure*}

A number of RFI mitigation methods \citep[e.g.,][]{Fridman2001,Ford2014,Baek2015,Wang2021,Zhang2021rfi,hu2021} could be used at different stages in the data acquisition process. After collecting astronomical data, however, the only thing that astronomers can do for RFI Mitigation is to flag and eliminate the effects of known and unknown sources of RFI. In the FAST astronomical spectral data, flagging RFI is seriously affected by the baseline distribution, due to that the baseline is not flat enough. Therefore, we have to firstly choose an efficient method, which is the ArPLS-based method \citep{Baek2015}, to remove the baseline, and then we could flag different types of RFI by experientially setting a credible threshold for each spectrum.

\subsection{The ArPLS for baseline correction}

For FAST spectrum data, We propose a baseline fitting method based on an Asymmetrically Reweighted Penalized Least Squares algorithm \citep[ArPLS;][]{Baek2015}. The baseline correction methods based on penalized least squares have been successfully applied to various spectral analysis \citep[e.g.,][]{Zeng2021,Zhang2021}. The methods could change a parameter ``weight'' iteratively to estimate a baseline. If a signal is below a previously fitted baseline, large weight is given. On the other hand, no weight or small weight is given when a signal is above a fitted baseline as it could be assumed to be a part of spectral peak. A preset coefficient $\lambda$ will control the balance between fitness and smoothness (see FAST spectrum with baseline correction in Figure\,\ref{Fig:baseline_fit}). To decrease computing resources, we choose the fitted spectral channels with one out of every 100 channels. In this case, we experientially set $\lambda$ to be 100, a constant, which could obtain satisfactory results for all of the available FAST data. This method could estimate the noise level iteratively and adjust the weights correspondingly \citep{Baek2015}.

The penalized constraint in this method makes the baseline fitting more robust and accurate than traditional methods by mitigating the negative influences from instrumental response \citep{Baek2015}. After removing the estimated baseline using the ArPLS (see Figure\,\ref{Fig:baseline_fit}), several novel strategies were proposed for flagging different types of RFI (see Section\,\ref{sect:flag_rfi}).


\subsection{Flag different types of RFI}
\label{sect:flag_rfi}

The RFI-flagging method mentioned above mainly targets at the FAST drift scan observations for searching for extra-galactic point sources. In this work, the used test data are from a blind drift scan survey\footnote{\url{https://fastwww.china-vo.org/cms/article/122/}}. This project is blindly mapping the sky using the idle time of the FAST observation. The project numbers are N2020\_3 and N2021\_4, respectively. The observation dates for our used data\footnote{At the end of July 2021, a strong standing wave (>1.0\,K) with a period of $\sim$8\,MHz has been removed from hardware.} are from July\,29 to September\,11 in 2021. The observation time is around $\sim$300 hours for our used test data in total. Based on the characteristics of the RFI in FAST after baseline correction, we adopt several strategies to flag them as follows.

\subsubsection{Flag extremely strong RFI}

In FAST observations, there indeed exist many extremely strong and evident RFI, for example those from communication satellites and navigation satellites \citep{Wang2021} at around 1175, 1207, 1270, and 1480\,MHz (see the emission line in Fig.\,\ref{Fig:baseline_fit}). These strong RFI have a serious effect on the FAST data reduction. These RFI could be easily removed by setting a reasonable threshold in the baseline fitted spectrum. For example, we could remove the spectral channels above power = $2\times10^{11}$ (see the lower panel of Figure\,\ref{Fig:baseline_fit}). However, if we could not make sure whether they are RFI, we could skip this strategy.

\subsubsection{Flag long-lasting RFI}

In the FAST drift scan survey, generally these long-lasting (e.g., > 128 seconds in this work) emission should be RFI except the \HI line at around 1420.5\,MHz. The reason is that almost all the distant extragalactic galaxies are close to point source relatively to the beam size of FAST, leading to that the signals cannot be lasting for long time\footnote{In a tracking mode or towards the relatively extended sources, this strategy is not applicable.}. Therefore, we could firstly average some spectra (e.g., averaging the neighborhood 8 seconds spectra for polarized \texttt{XX} and \texttt{YY} data, respectively, in this test work) that are close in time sequence (aim to improve signal-to-noise ratio), and then set a reasonable threshold to remove this kind of RFI (see Figure\,\ref{Fig:rfi_time}). Those still existing strong emission (power > $0.5\times10^{11}$; see Figure\,\ref{Fig:rfi_pol}) cannot be lasting longer than 128 seconds, thus they are not flagged as such kind of RFI, for example the blob RFI at 1255, 1280, and 1380\,MHz. These blob RFI could be found using the strategy of flagging beam-combined RFI (see Section\,\ref{sect:flag_beam}).

\subsubsection{Flag polarized RFI}

For FAST observation, a dual polarization receiver is equipped in the frontend, thus we could get two different polarized (\texttt{XX} and \texttt{YY}) emission simultaneously \citep{Jiang2019,Jiang2020}. A key fact is that the signals from the extragalactic galaxies are almost non-polarized, but most of the terrestrial RFI are polarized. Therefore, we could firstly average some spectra (e.g., averaging the neighborhood 8 seconds spectra for polarized \texttt{XX} and \texttt{YY} data, respectively, in this test work) that are close in time sequence (aim to improve signal-to-noise ratio). With their discrepancy, we then set a reasonable threshold to flag those polarized RFI from the terrestrial (see Figure\,\ref{Fig:rfi_pol}). Those still existing strong emission (power > $0.5\times10^{11}$; see Figure\,\ref{Fig:rfi_pol}) after this strategy are non-polarized RFI.

\subsubsection{Flag beam-combined RFI}
\label{sect:flag_beam}

The FAST has a powerful 19-beam receiver, which could simultaneously cover 19 different beam-size areas \citep{Jiang2019,Jiang2020}. In FAST drift scan survey, almost all the distant extragalactic galaxies are close to point source relatively to the beam size of FAST. This leads to that the true extragalactic signals could not simultaneously go into any two of the 19-beam receiver in a high probability. However, the observed terrestrial RFI by all the 19 receivers should get the same emission, except the occasionally and extremely straight RFI in direction. Therefore, we assume that the received RFI have no discrepancy in different beam receivers. We could firstly average some spectra (e.g., averaging the neighborhood 8 seconds spectra for polarized \texttt{XX} and \texttt{YY} data, respectively, in this test work) that are close in time sequence (aim to improve signal-to-noise ratio). Based on this advantage, we could then set a reasonable threshold to flag the beam-combined RFI (see Figure\,\ref{Fig:rfi_beam}). After flagging the beam-combined RFI, there exist several strong RFI, indicating this strategy is effective. Especially, most of the blob RFI at 1255, 1280, and 1380\,MHz have been well flagged out using this strategy. Those still existing strong RFI in Figure\,\ref{Fig:rfi_beam} could be occasional RFI coming into the receivers.

\subsubsection{Remove all the flagged RFI}

The upper panel of Figure\,\ref{Fig:rfi_all} shows a FAST spectrum after removing all the flagged RFI. We can see that those obvious RFI, compared with the spectrum in Figure\,\ref{Fig:baseline_fit}, have been removed from the whole wideband. The 2048 seconds waterfall image in the lower panel of Figure\,\ref{Fig:rfi_all} shows that the spectral data become very flat already. However, there still exist several RFI unflagged, for example at 1170, 1210, 1260, and 1480\,MHz. Currently in this work, we only flagged the RFI above $3\sigma\approx324$\,mK. The existing RFI could be further flagged by lowering down the flagging threshold. This also means that the threshold is sensitive to the number of the found RFI. Thus, the flagging threshold could be determined by the expected sensitivity or other requirements for each observation.

\section{RFI statistics}
\label{sect:statis}

\begin{figure*}
\centering
\includegraphics[width=0.90\textwidth, angle=0]{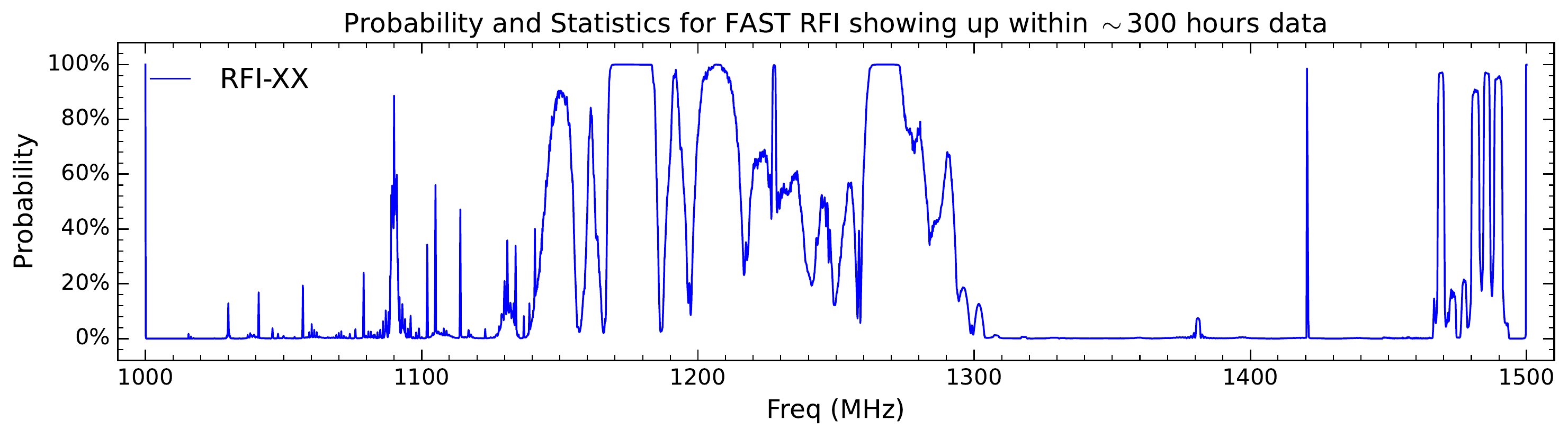}
\includegraphics[width=0.90\textwidth, angle=0]{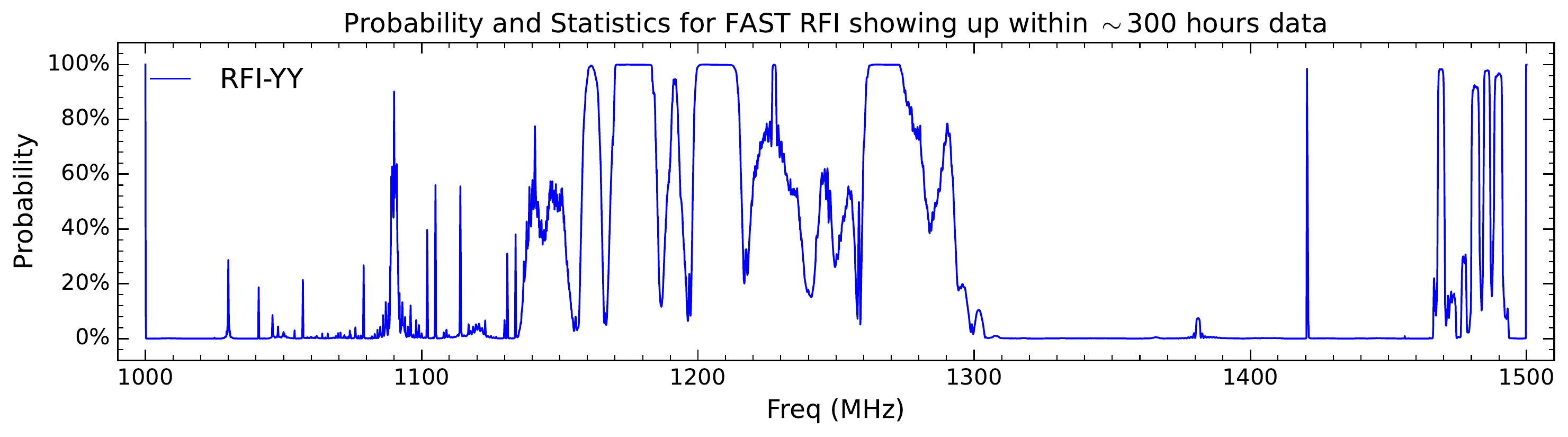}
\caption{Probabilities of RFI (\texttt{XX}, \texttt{YY}) showing up in FAST spectral observations. The corresponding data are listed in Table\,\ref{tab_probability}. The used statistical data have a span of 45 days, which amount to around $\sim$300 hours of observation time in total. Currently we only flagged the RFI above a threshold of $\sim$324\,mK. The emission at $\sim$1420.5\,MHz is \HI line from the Milky Way as background.}
\label{Fig:rfi_statis}
\end{figure*}

\begin{figure*}
\centering
\includegraphics[width=0.80\textwidth, angle=0]{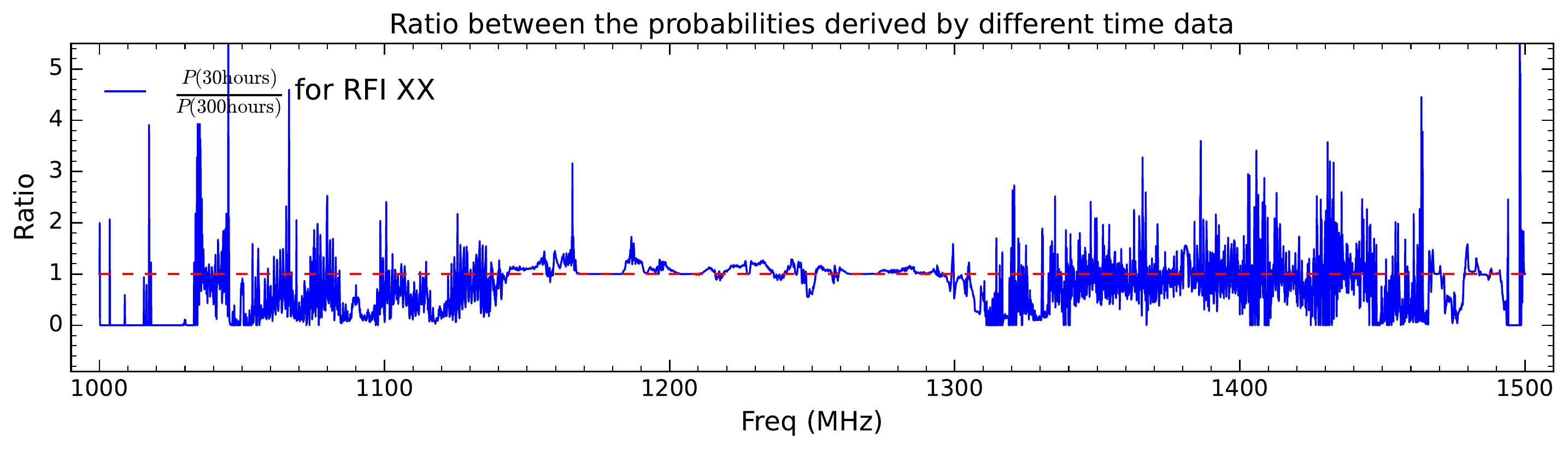}
\includegraphics[width=0.80\textwidth, angle=0]{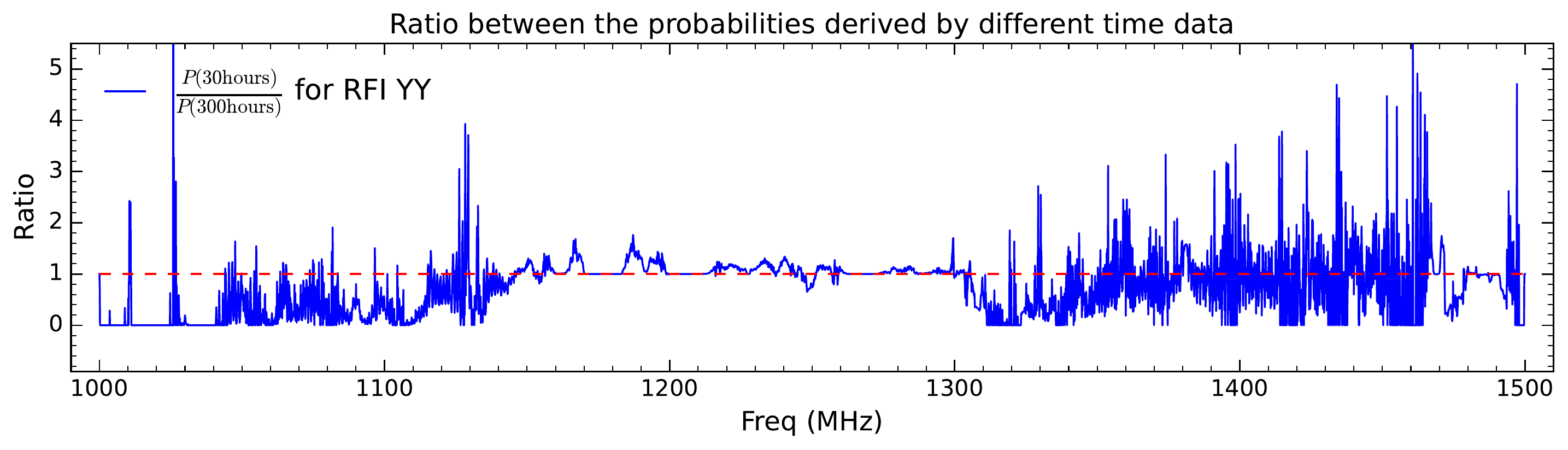}
\includegraphics[width=0.80\textwidth, angle=0]{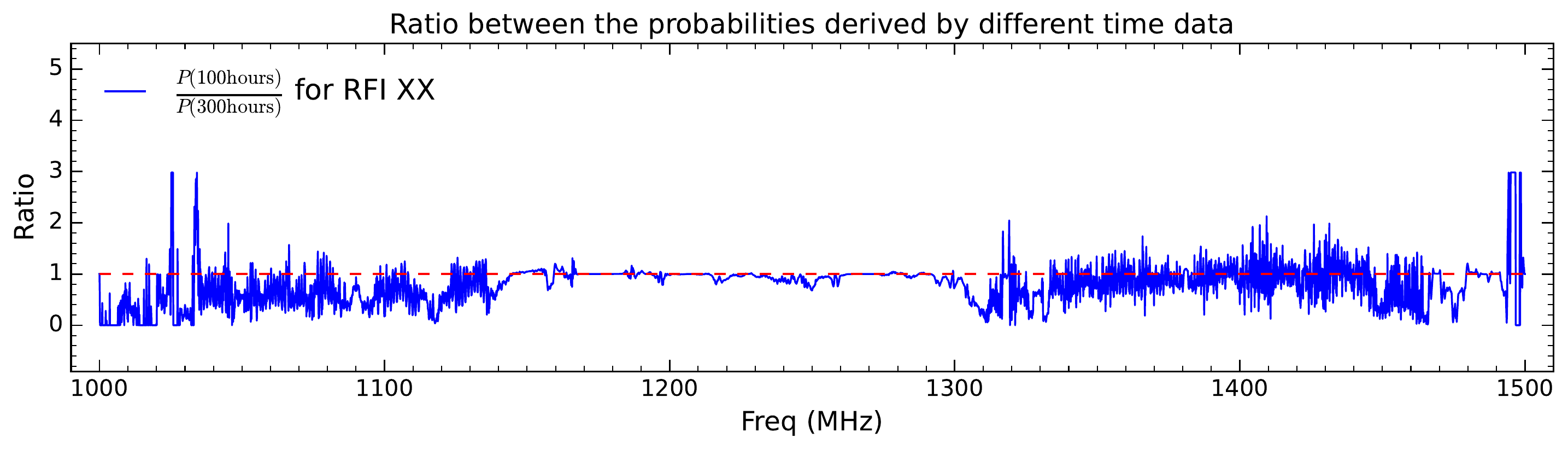}
\includegraphics[width=0.80\textwidth, angle=0]{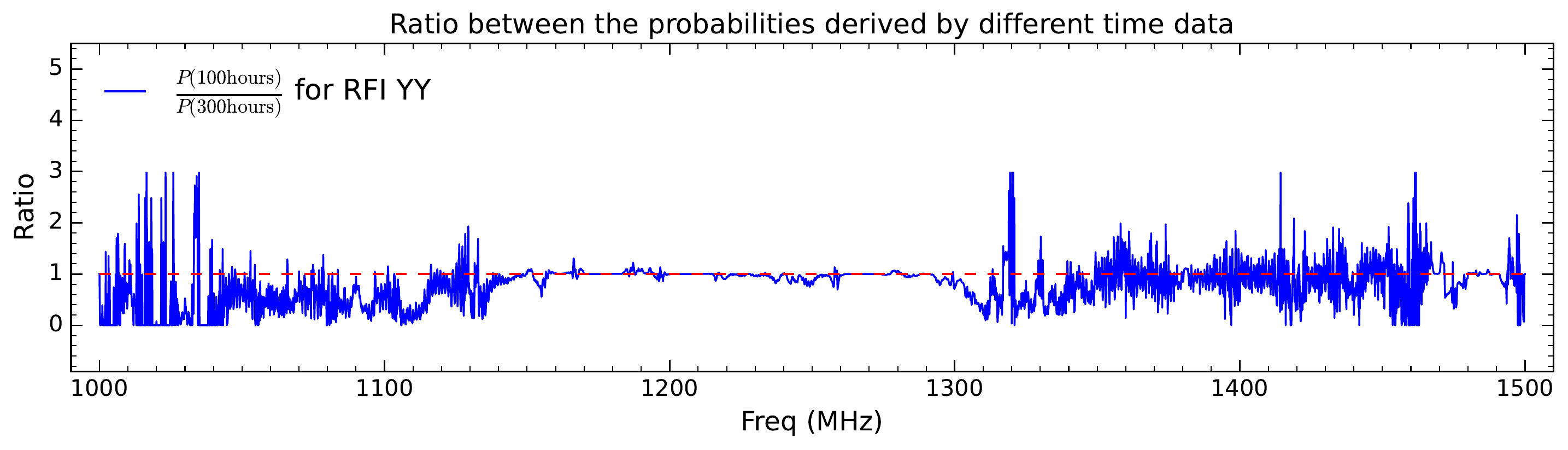}
\includegraphics[width=0.80\textwidth, angle=0]{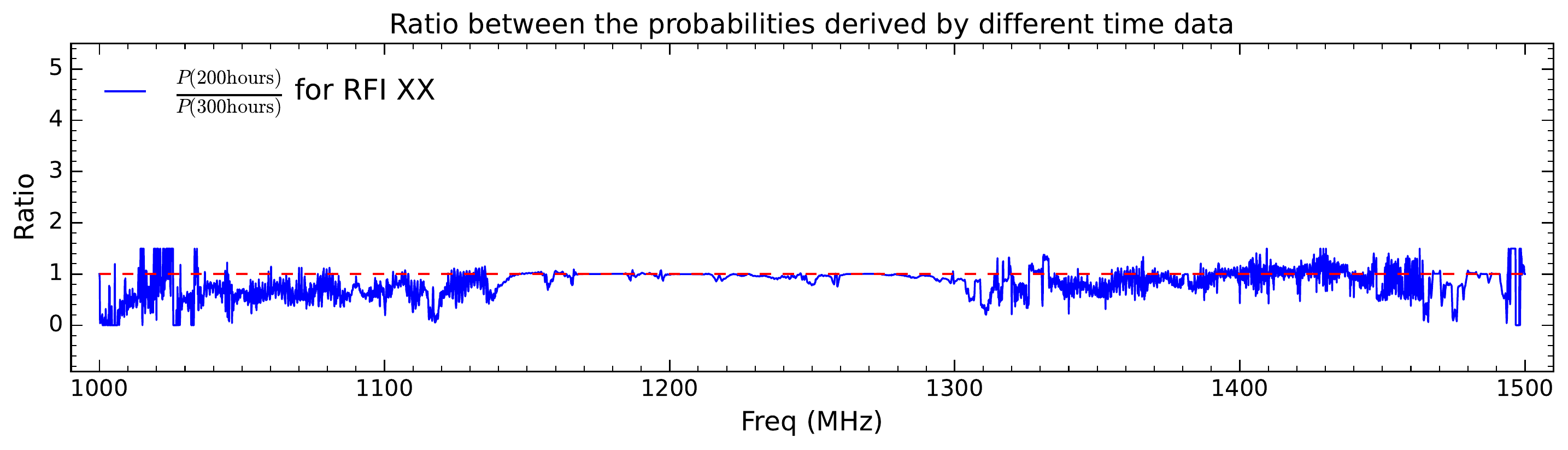}
\includegraphics[width=0.80\textwidth, angle=0]{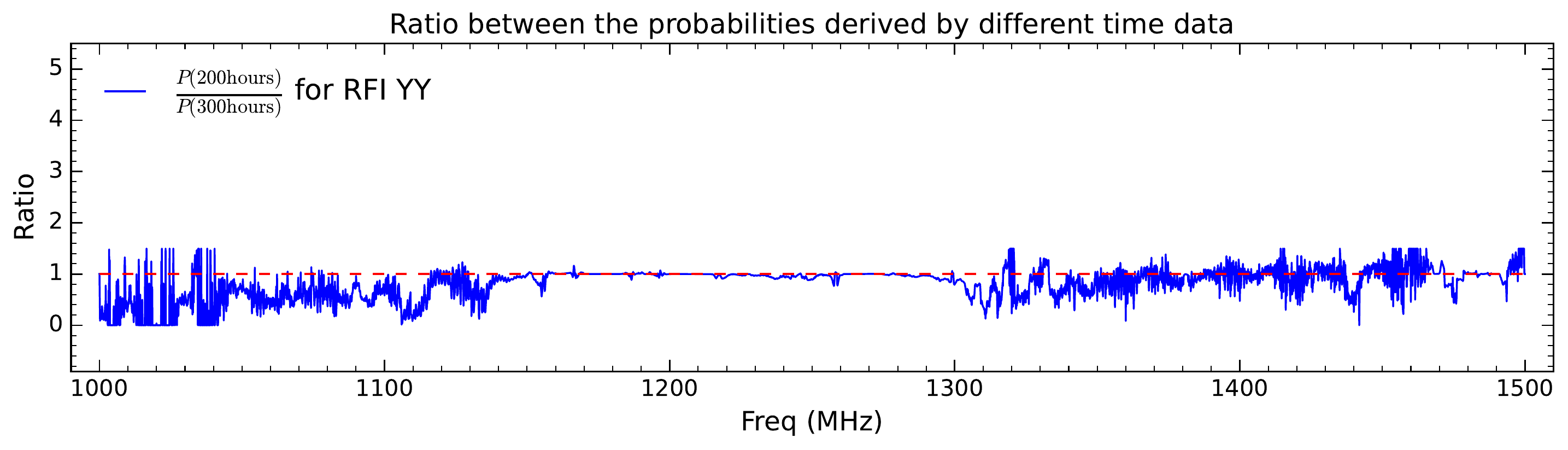}
\caption{Ratios between the flagged RFI (\texttt{XX}, \texttt{YY}) probabilities derived from 30, 100, 200, and 300\,hours. The RFI probabilities derived by 300\,hours observations are shown in Figure\,\ref{Fig:rfi_statis} and Table\,\ref{tab_probability}. The red lines indicate the values with ratio\,=\,1.}
\label{Fig:rfi_ratio}
\end{figure*}

\begin{figure*}
\centering
\includegraphics[width=0.80\textwidth, angle=0]{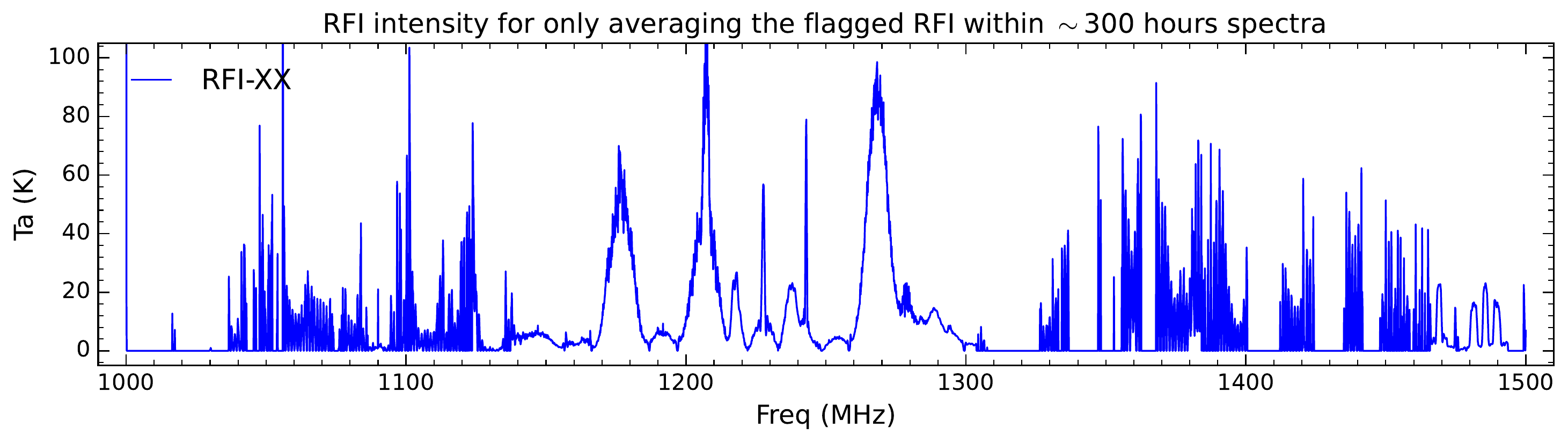}
\includegraphics[width=0.80\textwidth, angle=0]{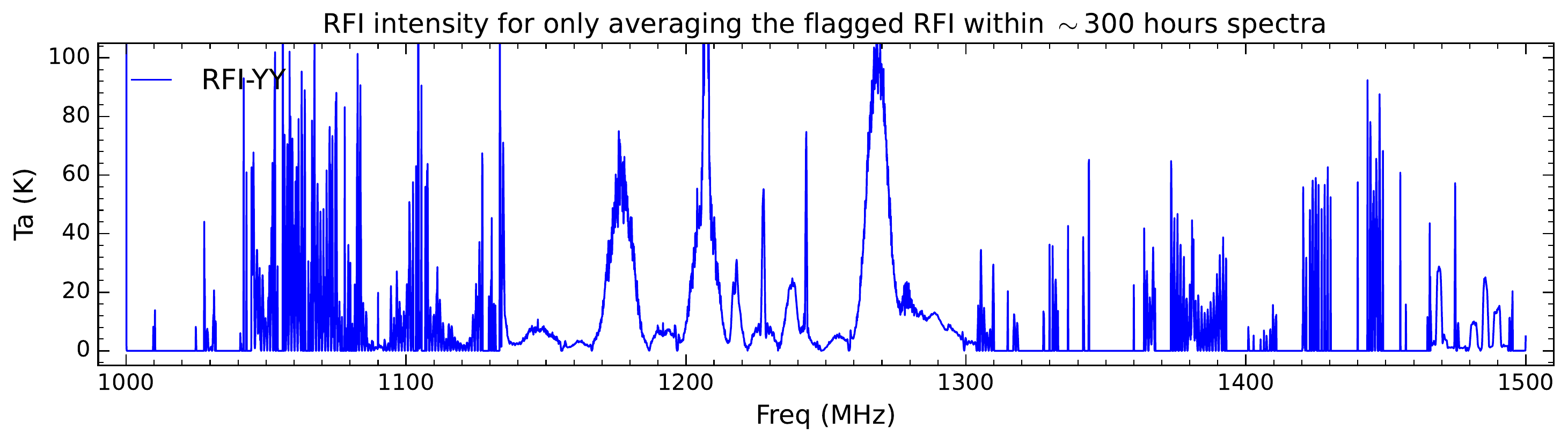}
\includegraphics[width=0.80\textwidth, angle=0]{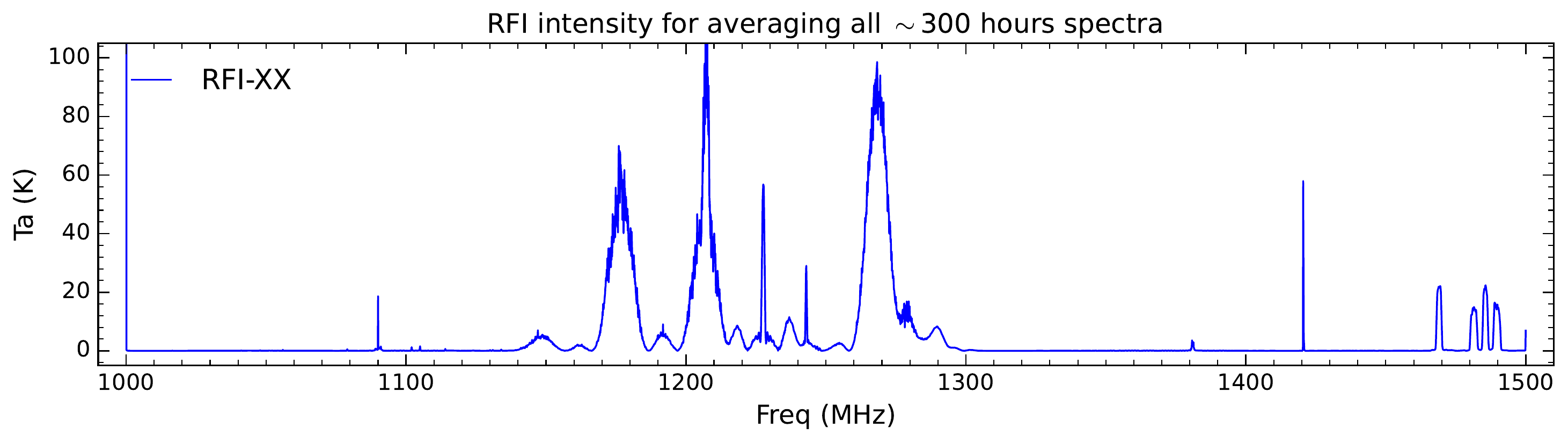}
\includegraphics[width=0.80\textwidth, angle=0]{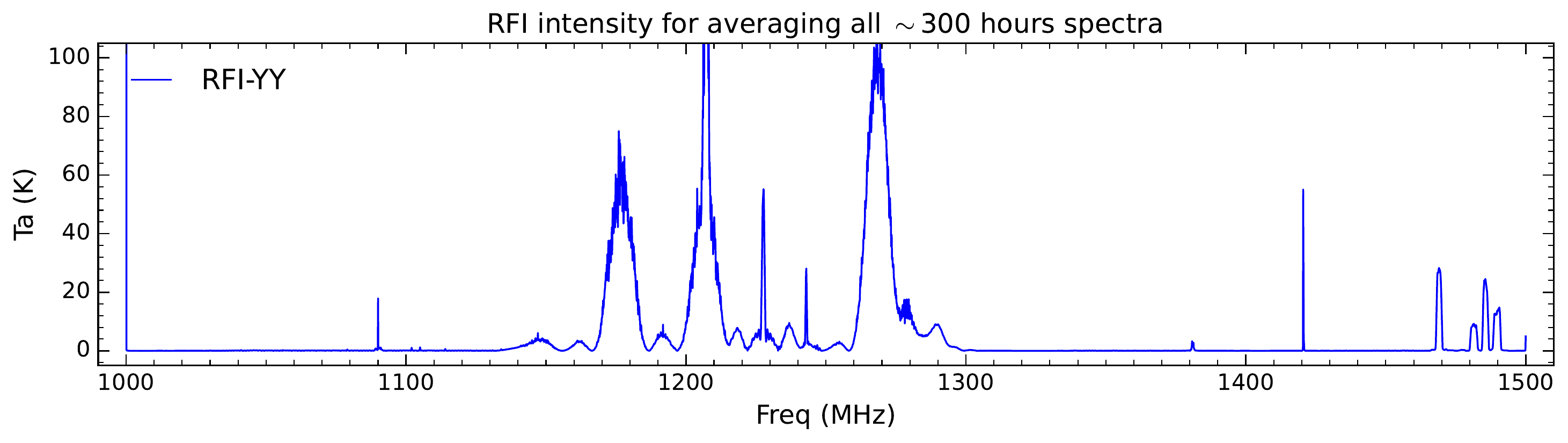}
\caption{Upper two panels: RFI (\texttt{XX}, \texttt{YY}) intensities from averaging the flagged RFI within $\sim$300 hours spectra. The intensities are listed in Table\,\ref{tab_probability} . Lower two panels:  RFI (\texttt{XX}, \texttt{YY}) intensities from averaging all $\sim$300 hours spectra.}
\label{Fig:rfi_intensity}
\end{figure*}

Based on the RFI flagging process, we find that there are indeed many RFI in the FAST spectral observation \citep{Jiang2020,Wang2021,Zhang2021rfi,hu2021}. We also find that some RFI are random and occasional, others are fixed at some frequencies\footnote{See also the test results from RFI monitoring in website: \url{https://fastwww.china-vo.org/cms/category/rfi_monitoring_en/}}. Since we already have a method which is able to flag the RFI above an expected threshold, it deserves to make a statistics for the probabilities of the RFI showing up in FAST spectral observations. The RFI statistical results have been shown in Figure\,\ref{Fig:rfi_statis} and Table\,\ref{tab_probability}. The used statistical data are the blind drift scan survey with project numbers are N2020\_3 and N2021\_4. All the data spanned 45 days from Jul.\,29 to Sep.\,11 in 2021. The total observation time reach to $\sim$300 hours for our used FAST data. Currently we only flagged the RFI above a threshold of $3\sigma\approx324$\,mK for each spectrum. The $\sim$300 hours statistical results would tell us which frequencies are relatively quiescent and which are noise. To do it, firstly we evenly bin the 500\,MHz wideband into 5000 channels\footnote{Off course it can be easily binned into more channels if needed.}. Each channel will have a bandwidth of 0.1\,MHz. With our RFI flagging procedure, we could find each RFI in every spectrum from long enough time observations. Then, we could compute the probability for each RFI showing up in each channel (0.1\,MHz bandwidth) and each polarization (\texttt{XX} or \texttt{YY}).

From Figure\,\ref{Fig:rfi_statis}, we find that the RFI with different polarizations (\texttt{XX} or \texttt{YY}) have obvious discrepancies at some frequencies. This further suggests that the terrestrial RFI are mostly polarized. In addition, the probabilities for the flagged RFI in most wideband are lower than 10\%, except the frequency range at 1155-1295\,MHz and 1465-1500\,MHz, where the RFI pollution is very serious with $P>60\%$. Even some RFI linewidths achieve to 20\,MHz. For the convenience of query, the full online table\footnote{\url{http://www.doi.org/10.11922/sciencedb.j00001.00313}} of Table\,\ref{tab_probability} lists the detailed RFI probabilities and antenna temperatures from 1000 to 1500\,MHz with a step of 0.1\,MHz. With the RFI statistical data, we are able to selectablely use some frequencies in the FAST observations.

Figure\,\ref{Fig:rfi_ratio} shows ratios between the flagged RFI (\texttt{XX}, \texttt{YY}) probabilities derived from 30, 100, 200, and 300\,hours. The ratios in different time tell us that: firstly, with increasing time the probability ratios become more and more stable, example ratio\,$=\frac{P(\rm200hours)}{P(\rm300hours)}<1.2$ at most frequencies; secondly, although $\sim$300 hours of observation time have been used for the statistics, every different observation may still meet different RFI possibly, especially for the small probability event (<\,0.1\%). This is because that some RFI are always occasionally appeared in some frequencies. Thus, the discrepancies will be much different, if the two observations are carried out in different days (or spanning many days).

In Figure\,\ref{Fig:rfi_intensity}, the upper two and lower two panels show RFI (XX, YY) intensities from averaging only the flagged RFI within $\sim$300 hours spectra, and those from averaging all $\sim$300 hours spectra, respectively. From the intensity discrepancies of RFI in Figure\,\ref{Fig:rfi_intensity}, we know that although some RFI are very bright, their probabilities showing up is very low, for example at around 1060, 1370, and 1450\,MHz. These frequencies are safe for observations in our opinion. Furthermore, in our strategies we only flag the RFI with intensity above some threshold. This will lead to that there is still some relatively weak RFI cannot be flagged below the used threshold. Therefore, current flagged RFI are still uncompleted, but the RFI probabilities showing up in $\sim$300 hours data could be a good reference for FAST users.

\section{Summary and future work}
\label{sect:summary}

In FAST spectral observations, the RFI always influences the identification of the interesting astronomical objects. Mitigating the RFI becomes an essential procedure in the data reduction. Thus, we provide a method to mitigate the RFI in FAST spectral observation and give a statistical result for the flagged RFI as follows.

Firstly, considering that the FAST spectra have a very wide bandwidth with 500\,MHz, but its baseline is not flat enough. According to the characteristics of FAST spectra, we propose to use the ArPLS algorithm to do baseline fitting. The test results show that the ArPLS has a really good performance in the baseline correction of the FAST spectral data reduction.

Secondly, we flag the RFI in FAST spectra in four strategies, which are flagging extremely strong RFI, flagging long-lasting RFI, flagging polarized RFI, and flagging beam-combined RFI, respectively. The test results (see Figure\,\ref{Fig:rfi_all}) show that all the RFI above the preset threshold could be removed from the whole wideband efficiently. Furthermore, this also means that if lowering down the threshold, more RFI could be found. Thus, the threshold is sensitive to the number of the found RFI. The derived RFI table is the most complete database currently for FAST.

\begin{table}[htp]
\caption{Probabilities and intensities for FAST RFI showing up in spectral observations.}
\label{tab_probability} 
\centering \small 
\setlength{\tabcolsep}{1.6mm}{
\begin{tabular}{c|rr|rr}
\hline \hline
Freq &  $P$(\texttt{XX}) & $T_{\rm a}(\texttt{XX})$ & $P$(\texttt{YY}) & $T_{\rm a}(\texttt{YY})$  \\
(MHz)   & (\%)  & (K)  & (\%)   & (K)     \\  
\hline
1420.0  &      0.3639  &      ---  &      0.0640  &      ---  \\
1420.1  &      0.2366  &      ---  &      0.0143  &      ---  \\
1420.2  &      0.2092  &      ---  &      0.0131  &      ---  \\
1420.3  &      0.6292  &      5.6910  &      0.4839  &      9.1579  \\
1420.4  &     58.1944  &      0.8797  &     58.0203  &      0.8658  \\
1420.5  &     98.6014  &     58.7972  &     98.5334  &     55.8846  \\
1420.6  &     78.8861  &     36.0836  &     78.9003  &     34.2117  \\
1420.7  &     67.0119  &      6.7575  &     66.9512  &      6.4334  \\
1420.8  &     42.7732  &      3.2709  &     42.6316  &      3.1254  \\
1420.9  &     15.2287  &      2.4960  &     15.2411  &      2.3321  \\
1421.0  &      6.5991  &      2.1241  &      6.5215  &      1.8191  \\
1421.1  &      1.9130  &      3.0898  &      1.9193  &      2.4175  \\
1421.2  &      0.5058  &      ---  &      0.3443  &      ---  \\
1421.3  &      0.3501  &      ---  &      0.2653  &      2.7763  \\
1421.4  &      0.2950  &      ---  &      0.1950  &      4.8645  \\
1421.5  &      0.2078  &      ---  &      0.1310  &     15.8759  \\
1421.6  &      0.1657  &      ---  &      0.1134  &     31.8494  \\
1421.7  &      0.1054  &     22.8050  &      0.0892  &      ---  \\
1421.8  &      0.1063  &     34.5292  &      0.0597  &      ---  \\
1421.9  &      0.1000  &     33.8378  &      0.0300  &      ---  \\
1422.0  &      0.1584  &     18.1468  &      0.0490  &      ---  \\
1422.1  &      0.1671  &      8.0035  &      0.0219  &      ---  \\
1422.2  &      0.1431  &      ---  &      0.0298  &      ---  \\
1422.3  &      0.1298  &      ---  &      0.0207  &      ---  \\
1422.4  &      0.1127  &      ---  &      0.0393  &      ---  \\
1422.5  &      0.1067  &      ---  &      0.0340  &      ---  \\
1422.6  &      0.1161  &      ---  &      0.0360  &      ---  \\
1422.7  &      0.0877  &      ---  &      0.0375  &      ---  \\
1422.8  &      0.1034  &     29.0777  &      0.0606  &      ---  \\
1422.9  &      0.1664  &     24.0432  &      0.1187  &     48.1041  \\
1423.0  &      0.1422  &     31.1379  &      0.1106  &     36.8297  \\
1423.1  &      0.1241  &     28.3946  &      0.0324  &      ---  \\
1423.2  &      0.1733  &      9.9554  &      0.0288  &      ---  \\
1423.3  &      0.1266  &      ---  &      0.0325  &      ---  \\
1423.4  &      0.1281  &      ---  &      0.0356  &      ---  \\
1423.5  &      0.1202  &      ---  &      0.0466  &      ---  \\
1423.6  &      0.1178  &      ---  &      0.0665  &      ---  \\
1423.7  &      0.0774  &      ---  &      0.0730  &      ---  \\
1423.8  &      0.0611  &      ---  &      0.1016  &     53.3374  \\
1423.9  &      0.0692  &      ---  &      0.1031  &     58.1402  \\
1424.0  &      0.0752  &      ---  &      0.1242  &     52.7798  \\
1424.1  &      0.1114  &     45.7388  &      0.0830  &      ---  \\
1424.2  &      0.0429  &      ---  &      0.0713  &      ---  \\
1424.3  &      0.1022  &     22.6616  &      0.0802  &      ---  \\
1424.4  &      0.1266  &      ---  &      0.1092  &      ---  \\
1424.5  &      0.1200  &      ---  &      0.0955  &      ---  \\
1424.6  &      0.0703  &      ---  &      0.1044  &      ---  \\
1424.7  &      0.0805  &      ---  &      0.1136  &     19.3083  \\
1424.8  &      0.0586  &      ---  &      0.1126  &     36.5678  \\
1424.9  &      0.0459  &      ---  &      0.1214  &     50.4057  \\
1425.0  &      0.0633  &      ---  &      0.1177  &     59.0533  \\
\hline
\end{tabular}}
\begin{flushleft}
\normalsize
\textbf{Notes.} In table here, we only list the RFI data with frequencies between 1420.0 and 1425.0\,MHz. The full table (\url{http://www.doi.org/10.11922/sciencedb.j00001.00313}) has a frequency range from 1000.0 to 1500.0\,MHz with a step of 0.1\,MHz and can be downloaded online. The flagged RFI are located above a threshold of $\sim$324\,mK. $P$(\texttt{XX}) and $P$(\texttt{YY}) are the probabilities of polarized \texttt{XX} and \texttt{YY} RFI showing up, respectively. $T_{\rm a}(\texttt{XX})$ and $T_{\rm a}(\texttt{YY})$ are estimated from a mean value only for the flagged RFI during $\sim$300 hours observations, but the $T_{\rm a}(\texttt{XX})$ and $T_{\rm a}(\texttt{YY})$ are for reference only. The emission at $\sim$1420.5\,MHz is \HI line from the Milky Way as background.
\end{flushleft}
\end{table}

Thirdly, we make a statistics for the probability and intensity of the RFI (including the polarized \texttt{XX} and \texttt{YY}) showing up in FAST spectral observations (see Figure\,\ref{Fig:rfi_statis}, and Table\,\ref{tab_probability}). The statistical results could tell us which frequencies are relatively quiescent. With such statistical data, the astronomers could try to avoid using such frequencies in the spectral observations, and our FAST staff will also try to mitigate the RFI.

In future work, firstly, the standing wave, for example the 1\,MHz wide standing wave, also seriously interferes the FAST observations, especially in extragalaxy survey. If this standing wave could be removed, the sensitivity will be improved several times. We are trying to remove it. Secondly, currently we don't know whether the RFI showing up is regular or not in different azimuth and zenith at different time (day or night). With our introduced strategies and using the huge FAST archival data, we will make a detailed statistical work to answer this question and tell FAST users how to avoid the terrible RFI. Thirdly, we are considering to equip the script into the FAST observing or data reduction system, to real-time monitor RFI, and to provide FAST users for an RFI data base for reference.

\section*{Acknowledgements}
\addcontentsline{toc}{section}{Acknowledgements}

This work is supported by the National Key R\&D Program of China (2018YFE0202900). FAST is a Chinese national mega-science facility, operated by the National Astronomical Observatories of Chinese Academy of Sciences (NAOC). We acknowledge support by the NAOC Nebula Talents Program and the Cultivation Project for FAST Scientific Payoff and Research Achievement of CAMS-CAS. We also wish to thank the anonymous referee for comments that improved the clarity of the paper.

\bibliographystyle{raa}
\bibliography{ms2021-0304}

\clearpage

\end{document}